\documentclass[usenatbib]{mnras}
\pdfoutput=1
\usepackage{amsmath}
\usepackage{amsfonts}
\usepackage{amssymb}
\usepackage{times}
\usepackage{graphicx}
\usepackage[T1]{fontenc}
\usepackage{aecompl}
\usepackage{verbatim}

\title[X-rays from Green Pea Analogs]{X-rays from Green Pea Analogs}
\author[M. Brorby, P. Kaaret]{M. Brorby$^{1}$\thanks{E-mail:
matthew-brorby@uiowa.edu}, P. Kaaret$^{1}$\\
$^{1}$Department of Physics and Astronomy, University of Iowa, Iowa City, IA 52242\\
}
\begin{document}

\pagerange{\pageref{firstpage}--\pageref{lastpage}} \pubyear{2017}

\maketitle

\label{firstpage}

\begin{abstract}
X-ray observations of two metal-deficient luminous compact galaxies (LCG) (SHOC~486 and SDSS J084220.94+115000.2) with properties similar to the so-called Green Pea galaxies were obtained using the {\emph{Chandra X-ray Observatory}}. Green Pea galaxies are relatively small, compact (a few kpc across) galaxies that get their green color from strong [OIII]$\lambda$5007\AA\ emission, an indicator of intense, recent star formation. These two galaxies were predicted to have the highest observed count rates, using the X-ray luminosity -- star formation rate ($L_X$--SFR) relation for X-ray binaries, from a statistically complete sample drawn from optical criteria. We determine the X-ray luminosity relative to star-formation rate and metallicity for these two galaxies. Neither exhibit any evidence of active galactic nuclei and we suspect the X-ray emission originates from unresolved populations of high mass X-ray binaries. We discuss the $L_X$--SFR--metallicity plane for star-forming galaxies and show that the two LCGs are consistent with the prediction of this relation. This is the first detection of Green Pea analogs in X-rays.
\end{abstract}


\section{Introduction}\label{sect:intro}
\begin{figure}
\centering
\includegraphics[width=0.5\textwidth]{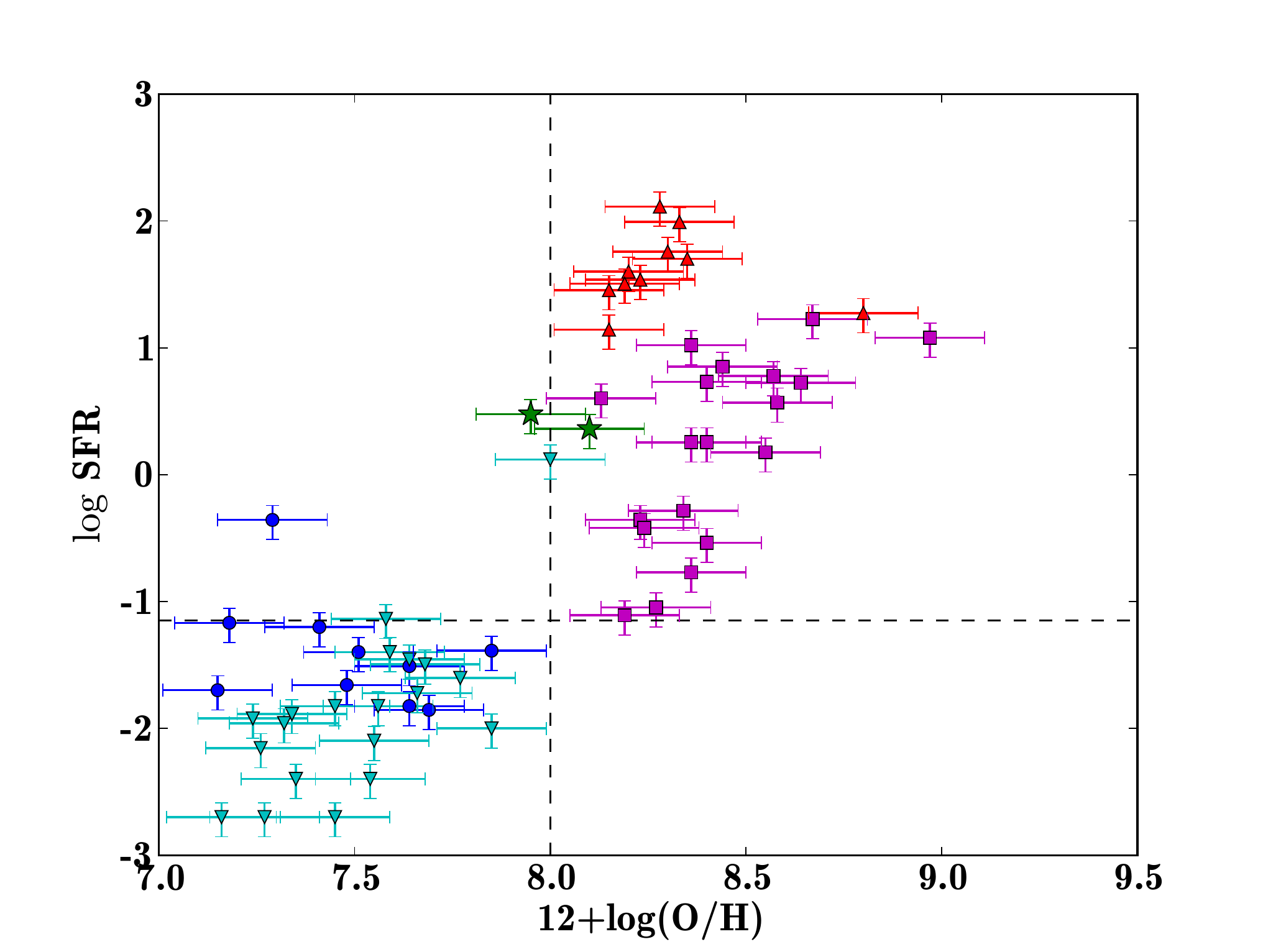}
\caption{Plot of SFR--metallicity from Brorby+2016 with the addition of the two Green Pea analogs (green stars). Note that the Green Pea analogs occupy the same parameter space as the larger spirals and irregular galaxies (magenta squares). Legend is given in Figure~\ref{fig:plane}.}\label{fig:sfrmet}
\end{figure}

In the early Universe $(z>6)$, during the time of reionization, stars and galaxies likely formed from pristine neutral gas, giving these objects very low metallicities~\citep{Loeb2010}.
In recent years, X-ray observations of star-forming galaxies have shown a metallicity dependence in the number of bright X-ray sources produced~\citep{Prestwich2013,Brorby2014} and the X-ray luminosity-star formation rate ($L_X$--SFR) relation~\citep{Basu-Zych2013a,Basu-Zych2013,Brorby2016}. This has implications on heating and reionization in the early Universe~\citep[e.g.,][]{Mirabel2011,McQuinn2012,Mesinger2013,Fragos2013a}. 
Recently, \cite{Brorby2016} have shown that the metallicity dependence of the $L_X$--SFR relation can be described as a plane in the 3-dimensional parameters space ($L_X$--SFR--metallicity plane). This relation was constructed from samples of three distinct populations of galaxies occupying different regions of the SFR--metallicity space~(Figure~\ref{fig:sfrmet}). The galaxy types were: (1) a group of normal spiral and irregular galaxies with near-solar metallicities from \cite{Mineo2012a}, (2) a sample of Lyman break analogs which had sub-solar metallicities and properties similar to the Lyman Break galaxies~\citep{Basu-Zych2013,Brorby2016}, (3) and a sample of extremely low-metallicity $(<0.1\, Z_\odot)$ blue compact dwarf galaxies~\citep{Brorby2014}, which are the best local proxies to the first galaxies in the Universe~\citep{Kunth2000}.
By studying compact, star-forming dwarf galaxies whose SFR and metallicity put them in the same parameter space as the larger star-forming spirals and irregulars~(Figure~\ref{fig:sfrmet}), we can begin to test the robustness of the $L_X$--SFR--metallicity plane. A recently identified galaxy type that may fill this role are the so-called Green Pea galaxies.

A new population of compact starburst galaxies was discovered in 2009 by the volunteers participating in the Galaxy Zoo Project~\citep{Cardamone2009}. Galaxies within this population were given the name Green Peas due to their compact size and green appearance in the \textit{gri} composite images from SDSS. The green color is caused by a strong [OIII]$\lambda$5007\AA\ emission line, an indicator of recent star formation, appearing in the $r$ band. However, this restricts Green Peas to the redshift range of $z=0.112-0.360$. Follow-up studies by \cite{Amorin2010} and \cite{Izotov2011} found that Green Peas are a special-case ($0.112<z<0.360$) of luminous compact galaxies (LCG) with strong [OIII] emission, spanning a much broader range of redshifts, $z=0.02-0.63$. Using the sample of LCGs found by \cite{Izotov2011}, we observed two of the most promising candidates with the \textit{Chandra X-ray Observatory} and the \textit{Hubble Space Telescope} in order to determine whether there is enhanced X-ray emission with respect to SFR, and if such emission is associated with an HII star-forming region.

In Section~\ref{sect:sample}, we describe the selection process used on optical data to find the best candidates for X-ray observations. The X-ray observations, metallicity measurements, and previous multiwavelength studies of these objects are discussed in Section~\ref{sect:observations}. The procedure by which we determine the X-ray properties of the two galaxies is given in Section~\ref{sect:procedure}. Section~\ref{sect:results} presents our results and a discussion of the findings in the context of previous studies.

We assume cosmological parameters of $H_0 = 67.8$~km/s/Mpc and $\Omega_m = 0.308$ \citep{PlanckCollaboration2015} for all relevant calculations.

\section{Sample Selection}\label{sect:sample}

\begin{figure}
\centering
\includegraphics[width=0.49\textwidth]{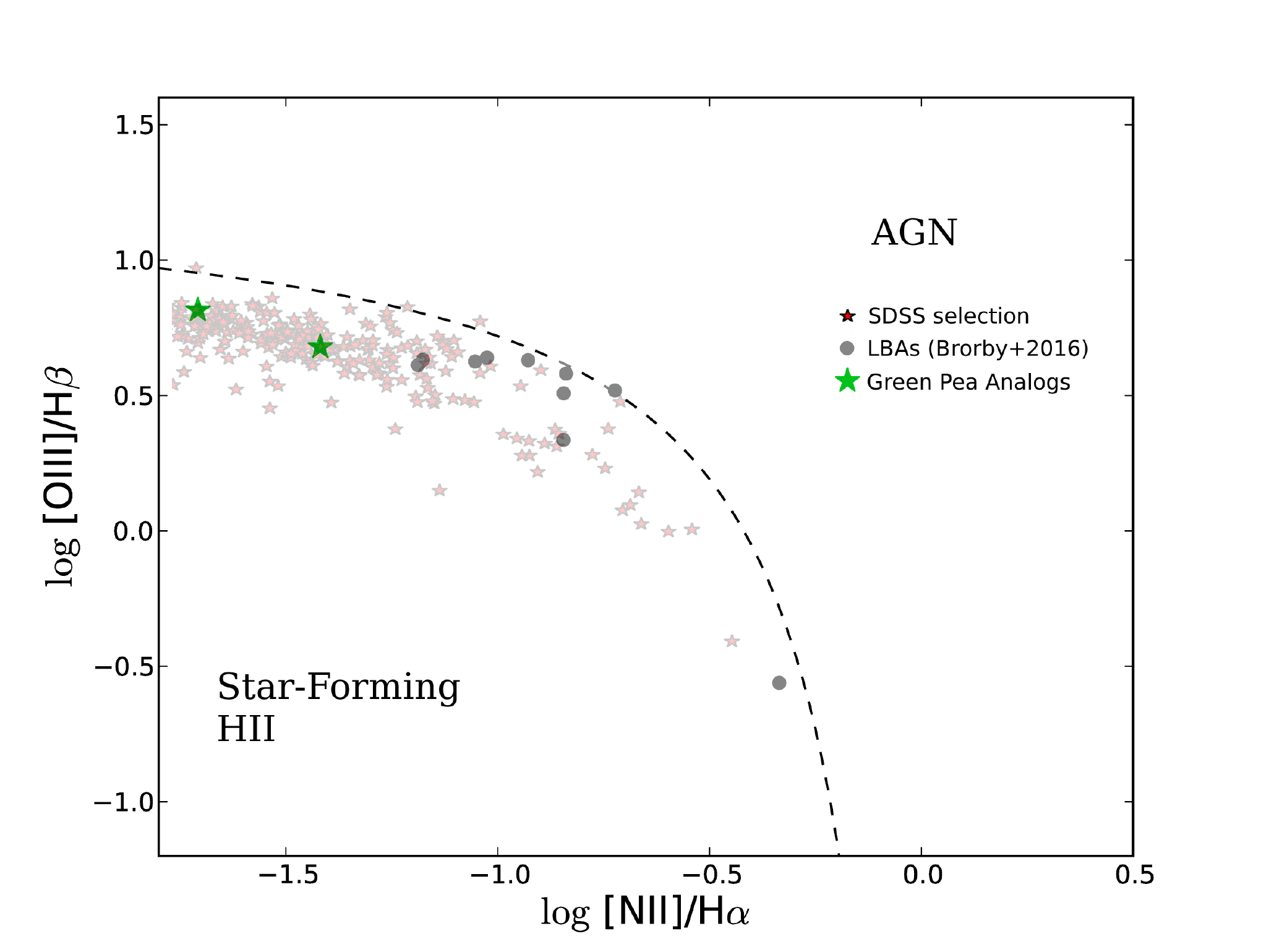}
\caption{BPT diagram. Plot of Lyman break analogs (grey circles), initial SDSS selection (faded red stars), and the two Green Pea analogs (green stars) from this paper. The BPT diagram uses SDSS emission line data to discriminate between clear AGN and star-forming, HII galaxies. All galaxies fall within the strict star-forming galaxy region of the diagram.}\label{fig:bpt}
\end{figure}

Our sample is selected from the Sloan Digital Sky Survey Data Release 12~(SDSS DR12)\footnote{\url{http://www.sdss.org}} and derived quantities from MPA-Garching and \cite{Brinchmann2008} using the selection criteria for LCGs as described by \cite{Izotov2011}. This criteria includes strong H$\beta$ emission, which is indicative of young starbursts~\citep{Izotov2011}. The galaxy spectra are also selected such that they exhibit well-detected [OIII]$\lambda$4363\AA\ emission with flux errors less than 50 per cent of the line flux, which allows for an accurate abundance measure using the `direct method' based on the determination of the electron temperature \citep[T$_e$-method; e.g.,][]{Pagel1992}. The galaxies are also required to be compact ($\sim$ 1-2 kpc in diameter). In order to distinguish our sample from the previously studied blue compact dwarf galaxies (BCDs)~\citep{Kaaret2011, Prestwich2013, Brorby2014}, which have similar stellar masses, physical sizes, compactness, and gas mass fractions, but have SFRs on the order of $10^{-2}~M_\odot$ yr$^{-1}$, we applied a lower limit for the SFR of $1~M_\odot$ yr$^{-1}$. 
We used [OIII]$\lambda 5007$, [NII]$\lambda 6584$, H$\alpha$, and H$\beta$ to construct a Baldwin-Phillips-Terlevich (BPT) diagram~\citep{Baldwin1981} as a diagnostic to distinguish star-forming galaxies from those hosting active galactic nuclei (AGN). We selected only those galaxies that fell on the star-forming side of the strict star-forming/AGN cut (dashed line, Figure~\ref{fig:bpt}). The SDSS spectra were also examined to ensure that there were no strong emission lines that are associated with Seyfert galaxies. We limited the sample of LCGs to have an abundance such that {12+log(O/H)$<$8.1}, which places them in the range where we expected an enhancement in the X-ray luminosity, based on the results from other low-metallicity galaxies \citep{Basu-Zych2013,Douna2015,Brorby2016}. To maximize the efficient use of observation time with Chandra, we imposed an upper limit on redshift such that $z<0.1$.

From this sample, we selected the LCGs that required the least observing time with \emph{Chandra}, based on conservative estimates of X-ray emission from HMXBs using the $L_X$--SFR~\citep{Mineo2012a} and redshift-dependent distance estimates. To convert the luminosity estimates to count rates, we used the WebPIMMS\footnote{\url{http://cxc.cfa.harvard.edu/toolkit/pimms.jsp}} Chandra proposal tool, assuming an absorbed power law spectrum with photon index $\Gamma = 1.7$ and the Galactic absorbing column density was determined along the line of sight using the Colden\footnote{\url{http://cxc.cfa.harvard.edu/toolkit/colden.jsp}} proposal tool. The observations times were determined by considering the total number of counts needed to obtain a significance threshold of $10^{-5}$ above background. We were awarded joint \emph{Chandra}/\emph{HST} observations of the two candidates with the shortest required exposure times: SHOC~486 and SDSS J084220.94+115000.2. The remaining sample candidates required observation times $\gtrsim 100$ks.

\cite{Shirazi2012} classified, with certainty, that SHOC 486 is a Wolf-Rayet galaxy -- a galaxy that exhibits evidence of Wolf-Rayet stars in its integrated spectra. WR galaxies are typically identified by a broad He~II~$\lambda4686$~\AA\ feature in the integrated emission spectra.
\cite{Brinchmann2008} classify both of these Green Pea analogs as clearly Wolf-Rayet (WR) galaxies. The selection criteria used by \cite{Brinchmann2008} to identify WR galaxies in SDSS are:
(1) spectrotype="galaxy" (2) EW$(\text{H}\beta) > 2$~\AA\ (3) S/N $>3$ in H$\beta$, [O{III}]$\lambda$5007, H$\alpha$, and [N{II}]$\lambda$6584. Additional selections are made to pick out WR galaxies, such as a broad component to He~{II}~$\lambda$4686 as well as N~{III}~$\lambda$4640 and other very clear WR features. Of the 20 galaxies within our candidate sample list that overlap with the \cite{Shirazi2012} sample, 16 galaxies did not show any clear WR features and were classified as non-WR galaxies. Thus, the two galaxies we observed may be evolutionarily distinct from a majority of the population that show no WR features. In all these galaxies, many \citep{Garnett1991,Thuan2005,Izotov2005} suggest that fast radiative shocks could be responsible for the nebular HeII~4686 emission.

\cite{Shirazi2012} examined a sample of SDSS galaxies with strong nebular HeII~4686 emission and discuss the possibility that the He$^+$ ionizing photons at low metallicities come from X-ray binaries. If true, it would indicate an increasing binary fraction with decreasing metallicity. From observations, it has been shown that the X-ray binary fraction does increases dramatically at low-metallicity \citep[e.g.,][]{Kaaret2011,Mapelli2012,Prestwich2013}.




\begin{table*}
\centering
\begin{minipage}{130mm}
\caption{LCG Photometric Results}\label{tab:results}
\begin{tabular}{llllll}
\hline
Target                              & SHOC 486           & SHOC 486 X-1  & J0842+1150        & J0842 X-1     & J0842 X-2     \\ \hline \hline
RA(J2000)                           & 14 48 05.30        & 14 48 05.36   & 08 42 20.90       & 08 42 20.93   & 08 42 20.74   \\
Dec. (J2000)                        & $-$01 10 57.7      & $-$01 10 57.8 & $+$11 50 00.2     & $+$11 50 00.2 & $+$11 50 01.5 \\
$n_\text{H}$ $(10^{20} \text{cm}^{-2})$   & 4.10        & $-$           & 4.15              & $-$           & $-$           \\
Metallicity $(12+\log(\text{O/H}))$ & 8.05               & $-$           & 8.09              & $-$           & $-$           \\
$D_L$ (Mpc)                         & 123.7              & $-$           & 133.1             & $-$           & $-$           \\
SFR $(M_\odot \text{yr}^{-1})$      & 5.4                & $-$           & 7.2               & $-$           & $-$           \\
Flux $(10^{-15} \text{erg~cm}^{-2}\text{~s}^{-1})$ & $7.7^{+4.4}_{-3.6}$ & $5.8^{+3.4}_{-2.5}$  & $17.0^{+3.6}_{-3.5}$ & $5.9^{+1.9}_{-1.6}$         & $3.4^{+1.7}_{-1.3}$          \\
$L_X$ $(10^{39} \text{erg~s}^{-1})$ & $14.1^{+8.0}_{-6.6}$ & $11^{+6}_{-5}$          & $36.0^{+7.6}_{-7.4}$ &    $12^{+4}_{-3}$ & $7^{+4}_{-3}$          \\
Obs. Time (ks)                      & 23.65              & 22.37         & 49.91             & 44.02         & 44.02         \\
$N_\gamma$ (photons)                & 16.6               & 11.6          & 81.8              & 32.3          & 12.7          \\
$N_b$ (photons)                     & 9.4                & 3.0           & 21.2              & 19.0          & 7.0          
\\ \hline
\end{tabular}
\\
\raggedright\textbf{Notes.} The table includes RA and DEC (J2000) of each galaxy and each X-ray source within the galaxies, metallicity (12+log(O/H))\citep{Brinchmann2008} $T_e$ method (uncertainties of $\pm 0.01$), luminosity distance in Mpc (from SDSS redshift)(old/new cosmology), star formation rate (SFR) from UV+IR, total X-ray luminosity ($L_X$) in the 0.5$-$8 keV band assuming a photon index of $\Gamma=1.7$, the observation time (ks), the number of photons from the source galaxy ($N_\gamma$), and the number of background photons ($N_b$) from the galaxy over the observation time. 
For SHOC~486, we used a D$_{25}$ of $10.68\arcsec\times 7.56\arcsec$ and angle $17.2\deg$. J0842+1150 has a D$_{25}$ of $10.43\arcsec\times 7.2\arcsec$ and angle $68.7\deg$.\\
\end{minipage}
\end{table*}

\section{Observations}\label{sect:observations}
\subsection{X-ray Observations}\label{sect:chandra_obs}
Our sample consists of two Green Pea analogs observed with the \emph{Chandra X-ray Observatory}. SHOC~486 is located at $\rmn{RA}(2000)=14^{\rmn{h}} 48^{\rmn{m}} 05\fs3$, $\rmn{Dec.}~(2000)=-01\degr 10\arcmin 57\farcs 7$ and was observed for $23.65$~ks on 2016-02-26. SDSS J084220.94+115000.2 (hereafter, J0842+1150) has $\rmn{RA}(2000)=08^{\rmn{h}} 42^{\rmn{m}} 20\fs9$, $\rmn{Dec.}~(2000)=+11\degr 50\arcmin 00\farcs 2$ and was observed on 2014-12-29 for a total of $49.91$~ks. These observations were obtained using the ACIS-S3 back-illuminated chip aboard Chandra. We reprocessed the level 1 event files using the latest version of CIAO~(4.8) and CALDB~(4.7.1). Table~\ref{tab:results} lists the names, coordinates, redshifts, and other derived values of the two galaxies.

We found that SHOC~486 exhibits a single, point-like X-ray source within a morphologically spherical galaxy. From HST images (Figure~\ref{fig:hst_xrayContours}), J0842+1150 appears to be an interacting galaxy consisting of two clumps of multiple star-forming regions, each with unresolved X-ray emission consistent with being a point source (see Figure~\ref{fig:xraySurfBright}). 

Using the $\log N - \log S$ curve of \cite{Lehmer2012} for background AGN, we calculate $N_\text{AGN}=0.01$, the number of AGN within the $D_{25}$ ellipse of either galaxy for flux values greater than $1\times 10^{-15}$~erg~cm$^{-2}$~s$^{-1}$, which is lower than the limiting fluxes for either observation. Thus, it is unlikely that these sources are coincident background AGN. For galaxies such as these, going through recent star formation, it is likely that the sources are dominated by bright X-ray binaries.

\subsection{Infrared and Ultraviolet SFR Indicator}\label{sect:sfr}
In order to compare the Green Pea observations with our previous studies, we use the UV+IR SFR indicator as outlined in \cite{Brorby2016}. The SFR indicator is calibrated to starburst galaxies and based on the indicator used by \cite{Mineo2012a} from the work of \cite{Bell2003,Hirashita2003,Iglesias-Paramo2004,IglesiasParamo2006}.
We use data from the WISE AllSky Survey \citep{wright2010}, which provided complete coverage of the sky in four infrared bands. Data from the archive were downloaded and the $22 \mu$m (WISE band 4) magnitudes were converted to monochromatic fluxes, as outlined by \cite{wright2010}. We used these derived fluxes in our SFR indicators.

A second component of the SFR indicators is the ultraviolet (UV) luminosity.
For each galaxy, we downloaded GALEX images from the archive. We use near UV (NUV, $\lambda_\text{avg} = 2312$\AA ) GALEX data to obtain UV luminosities.

For both the infrared and ultraviolet components of the SFR indicator, luminosities were determined from the net count rates within elliptical source regions using background annuli, as described for the X-ray data in Section~\ref{sect:procedure}. Conversions from count rate to flux can be found in \cite{Brorby2016} and references therein. The total SFR is given by
\begin{align}
\text{SFR}_\text{tot} &= \text{SFR}^0_\text{NUV} + \text{SFR}_\text{IR}\\
				      &= (1.2\times 10^{-43} L_\text{NUV,obs} [\text{erg s}^{-1}]) \notag\\
				      			\phantom{{}} & \phantom{={}} + (4.6\times 10^{-44} L_\text{IR} [\text{erg s}^{-1}]),
\end{align}
where $L_\text{NUV,obs}$ is the observed luminosity at 2312~\AA\ , uncorrected for dust attenuation, and $L_\text{IR}$ is the total IR luminosity ($8-1000$ micron) derived from WISE 22-$\mu$m luminosities. Our SFR estimates are presented in Table~\ref{tab:results}.

\subsection{Metallicity Measurements}\label{sect:metallicity}
\cite{Brinchmann2008} have used the well-detected [OIII]$\lambda 4363$\AA\ emission line to implement the direct $T_e$-method to determine oxygen abundances. They find oxygen abundances of $12+\log_{10}(\text{O/H}) = 8.05\pm0.01$ and $8.09\pm0.01$ for SHOC~486 and J0842+1150, respectively (see Table~\ref{tab:results}). We check these values by calculating the oxygen abundance using the O3N2-method. The O3N2-determined oxygen abundances for the two galaxies were calculated using line ratios from SDSS DR12 spectral data~\citep{abazajian2009} following the \cite{pettini2004} method. The O3N2-method requires emission line measures of [OIII]~$\lambda 5007$, [NII]~$\lambda6584$, H$\alpha$, and H$\beta$. The relation is given by $12+\log_{10}(\text{O/H}) = 8.73-0.32\times$O3N2, where O3N2 $= \log_{10}\{([\text{OIII}]/H\beta)/([\text{NII}]/H\alpha)\}$. The 68 per cent (95 per cent) confidence interval for this measure is $\pm 0.14\ (\pm0.25)$~dex.
The solar gas-phase metallicity is taken to be $12+\log_{10}(\text{O/H}) = 8.69$ \citep{AllendePrieto2001,asplund2004}, with an absolute metallicity of $Z_{\sun}=0.02$. Using the O3N2-method, we calculate metallicities of $7.95\pm0.14$ and $8.10\pm0.14$ for SHOC~486 and J0842, respectively. However, we adopt the \cite{Brinchmann2008} results, using the direct $T_e$-method, due to their lower uncertainties. 


\begin{figure}
\centering
\includegraphics[width=0.5\textwidth]{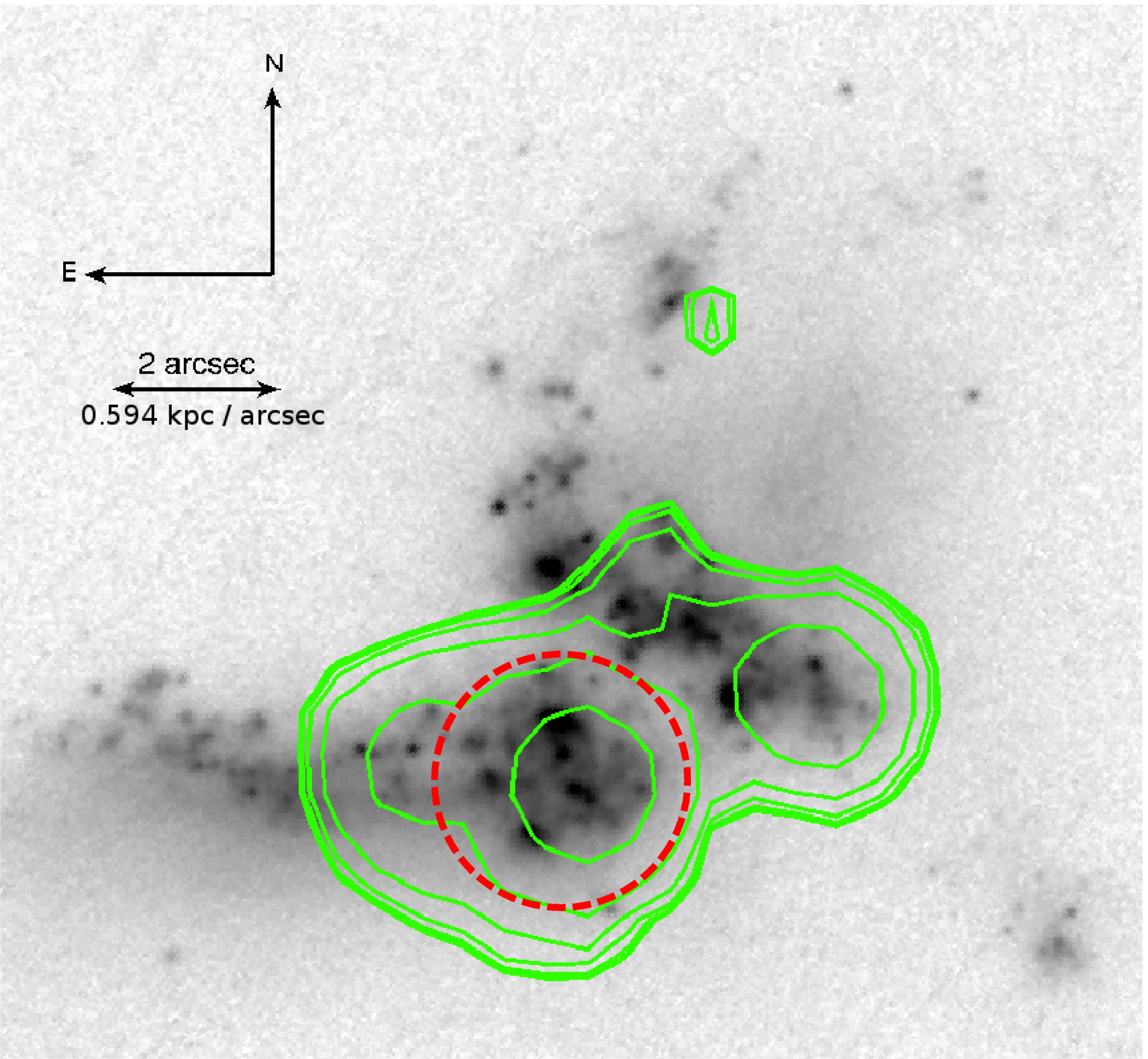}\\
\includegraphics[width=0.5\textwidth]{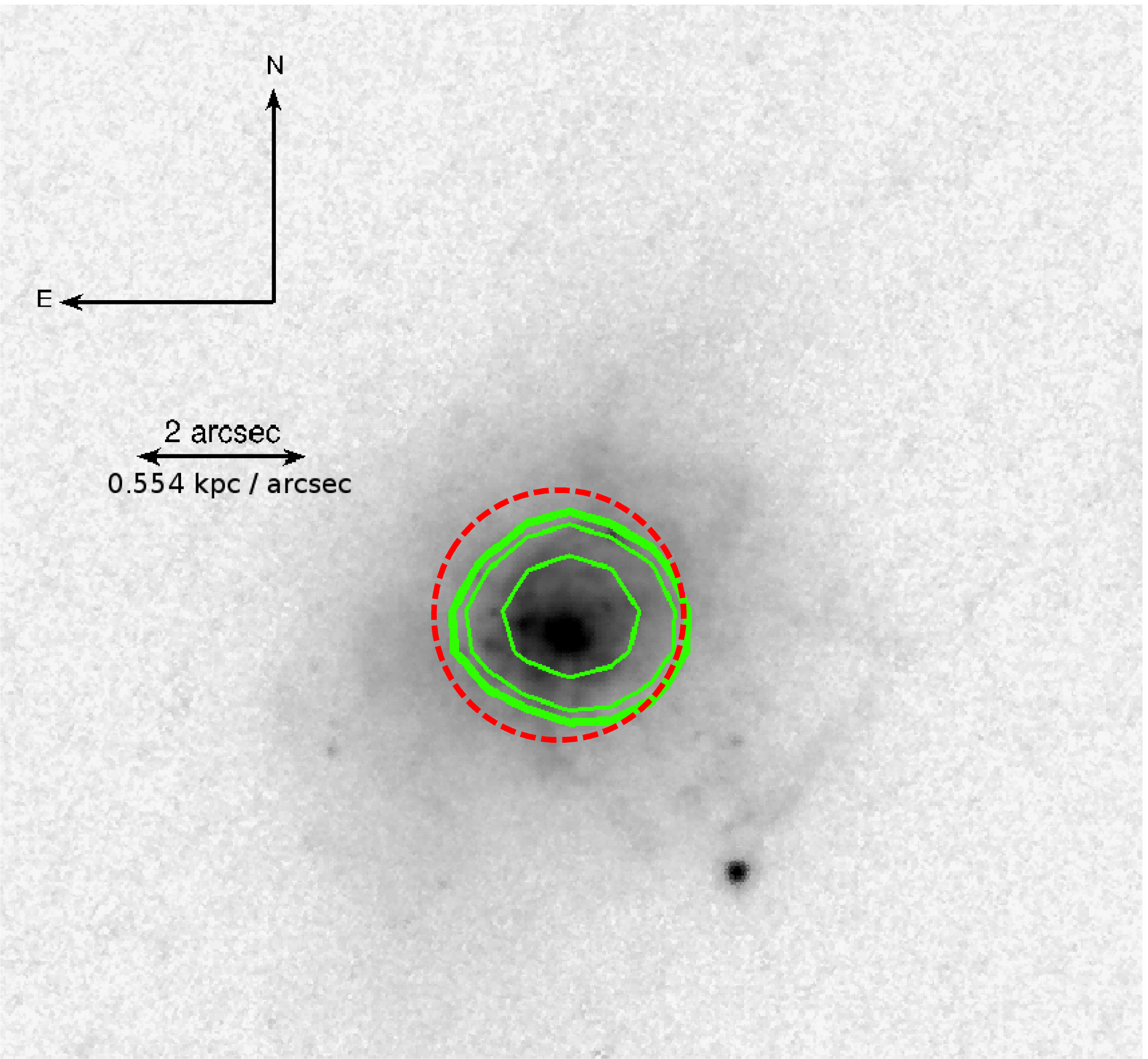}
\caption{SDSS spectral extraction regions, used for metallicity determinations, are shown as red, dashed lines (3$\arcsec$ diameter). Top: HST F606W image of J0842+1150 with X-ray contours (0.5-8 keV). The 2$\arcsec$ marker corresponds to 1.2~kpc. The small knot of apparent X-ray emission north of the dominant sources is not significantly detected. Bottom: HST F606W image of SHOC~486 with X-ray contours (0.5-8 keV). The 2$\arcsec$ marker corresponds to 1.1~kpc. The object to the south-west is classified as a foreground star in SDSS.}\label{fig:hst_xrayContours}
\end{figure}

\section{Procedure}\label{sect:procedure}
For both SHOC~486 and J0842+1150, we calculated the net number of counts in the $0.5-8$~keV energy range coming from within an ellipse, defined using the given dimensions in the HyperLeda\footnote{\url{http://leda.univ-lyon1.fr/}} database, centered on the galaxy. The Table~\ref{tab:results} notes contain the dimensions of these ellipses as well as their position angle. A larger source-free annulus, centered on the galaxy, was used for background estimation. Using these source and background regions, we ran the CIAO tool \texttt{srcflux} to calculate the unabsorbed flux in the $0.5-8$ keV energy band assuming an absorbed power law model. For each source, we determined the column density, $n_H$ by using the \texttt{prop\_colden} tool in CIAO (relevant values in Table~\ref{tab:results}). For the power law model, a photon index of $\Gamma=1.7$ was assumed. If one assumes $\Gamma=1.9$, the fluxes and luminosities are reduced by 10 per cent. The \texttt{srcflux} script provides unabsorbed, aperture-corrected net flux values based on the input model parameters and the source and background regions. This script uses up-to-date, energy-dependent values for the quantum efficiency and effective area across the detector. It is the preferred method over using PIMMS which only uses averaged estimates of these values.

For all sources within the two galaxies, we extract spectra using source regions that enclose a 90 per cent energy fraction based on the PSF-model for the observation. We fit these spectra in \texttt{XSPEC} using an absorbed power law (\texttt{wabs*pow}). For all sources, the column density, which determines the amount of absorption, could not be constrained and was set to the value used in Table~\ref{tab:results}. In fitting spectra with so few counts (Poisson data), one must abandon Gaussian statistics and instead use the \emph{C-statistic} for parameter estimation. And when testing goodness-of-fit, Pearson's chi-square test is preferred over the standard chi-square test, which again is only good for Gaussian data. Using the C-statistic, we determine the photon index, flux, and luminosity for each source. We find that the model parameters assumed in the photometric determination of luminosities are justified and give consistent results with our spectrally-determined values (see Table~\ref{tab:spectral}).

\begin{figure}
\centering
\includegraphics[width=0.49\textwidth]{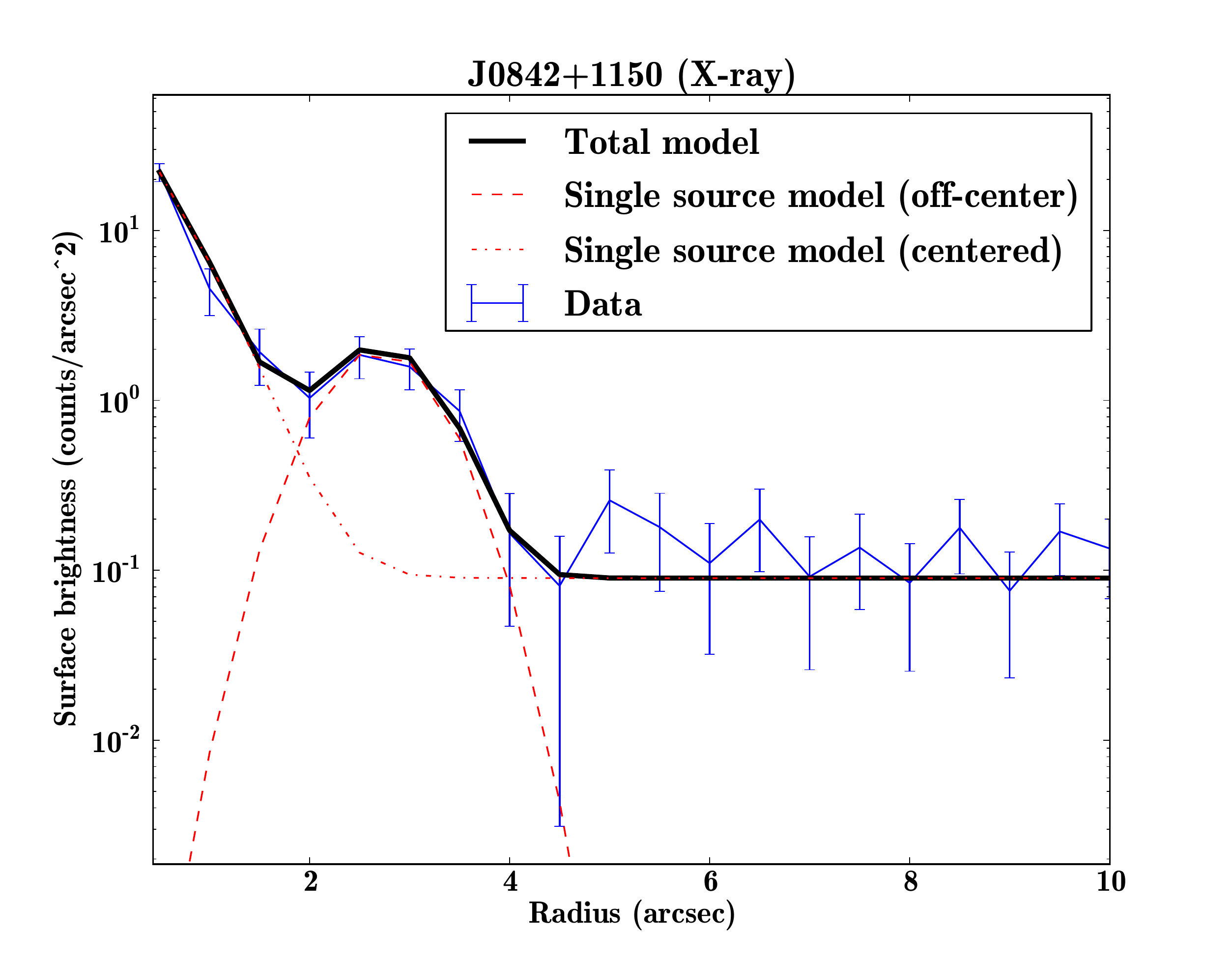}\\
\includegraphics[width=0.49\textwidth]{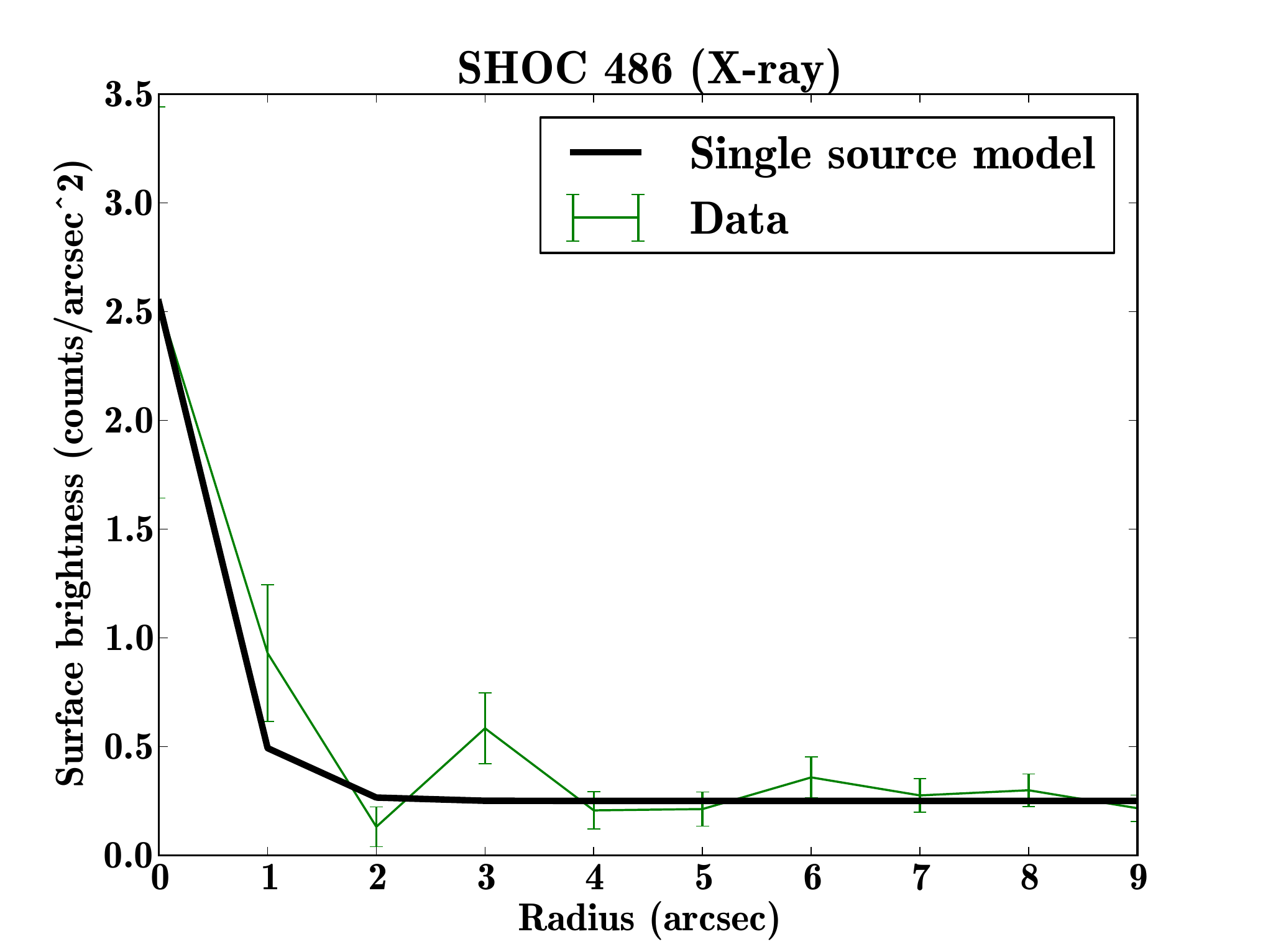}
\caption{These plots show the azimuthal average radial X-ray surface brightness profile, centered on the brightest X-ray source in each image. The X-ray sources (blue and green curves) in the two Green Pea analogs are consistent with being point sources (black curve). The underlying models (black curves) are PSF models from Chandra where we simply scale the model to match the central flux of the observed source(s). Top: X-ray surface brightness of the two X-ray sources in J0842+1150. The red, dashed curves are surface brightness models for the two X-ray sources, centered on the SDSS extraction region in Figure~\ref{fig:hst_xrayContours}. Bottom: X-ray surface brightness for the X-ray source in SHOC~486.}\label{fig:xraySurfBright}
\end{figure}

\section{Results and discussion}\label{sect:results}
\begin{figure*}
\centering
\includegraphics[width=0.75\textwidth]{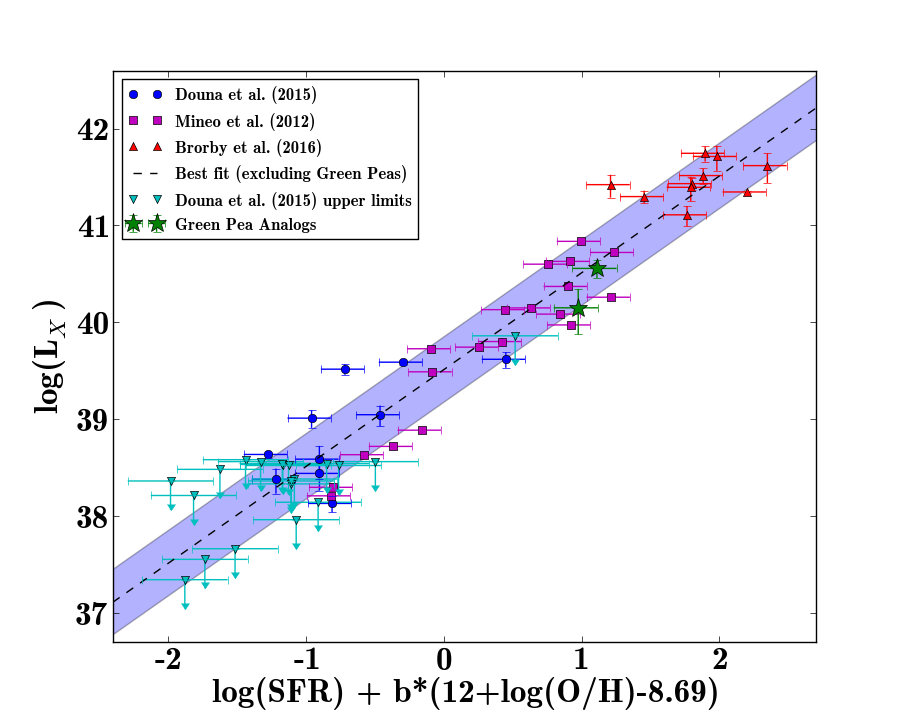}\\
\caption{Plot of $L_X$--SFR--metallicity from Brorby+2016 with the addition of the two Green Pea analogs (green stars). The dashed line represents a best-fit from Brorby+2016 with $b=-0.59\pm0.13$. We find that the Green Pea analogs are consistent with this relation. 
}\label{fig:plane}
\end{figure*}

We find X-ray luminosities of $L_X(0.5-8\text{keV})=1.4^{+0.8}_{-0.7}\times 10^{40}$~erg~s$^{-1}$ and $3.6^{+0.8}_{-0.7}\times 10^{40}$~erg~s$^{-1}$ within the entire defined ellipses for SHOC~486 and J0842+1150, respectively. Based on the associated SFR and metallicity for each galaxy, we use the $L_X$--SFR--metallicity relation of \cite{Brorby2016} to predict X-ray luminosities of $4.0^{+4.7}_{-2.1}\times 10^{40}$~erg~s$^{-1}$ and $5.0^{+6.0}_{-2.7}\times 10^{40}$~erg~s$^{-1}$, respectively. The predicted values have $1\sigma$ uncertainties based on the dispersion of the $L_X$--SFR--metallicity relation and the uncertainties in SFR and metallicity. The measured X-ray luminosities are in agreement with these predictions. 
In Figure~\ref{fig:plane}, we have plotted the two Green Pea analog galaxies from this paper onto the $L_X$--SFR--metallicity plane from \cite{Brorby2016} and find that they are consistent with the previous fit. Both fall within the 68~per~cent dispersion band shown in Figure~\ref{fig:plane}. These two galaxies show that this $L_X$--SFR--metallicity relation holds across a broad range of $L_X$ ($10^{37-42}$~erg~s$^{-1}$), SFR ($10^{-3}$--$10^2$~$M_\odot$~yr$^{-1}$), and metallicity ($0.03$--$1.6$~$Z_\odot$) as seen in Figure~\ref{fig:plane}. Across all these values, the dispersion is $\sigma = 0.34$~dex with a well-constrained normalization ($c=39.49\pm0.09$). Future work toward reducing the uncertainties in gas-phase metallicity by using the $T_e$-method may help to further constrain this planar relation. Large scatter in SFR indicators is still the largest contributor to the uncertainty.

From HST images (Figure~\ref{fig:hst_xrayContours}), we see that SHOC~486 is somewhat spherical in its morphology and contains a dense central region, from which the X-rays originate. To eliminate some of the diffuse X-ray emission, we reduce the X-ray extraction region to that of a point source with a 90~per~cent enclosed energy fraction. From this region, we measure a net flux of $5.8^{+3.4}_{-2.5}$ erg~cm$^{-2}$~s$^{-1}$ which corresponds to a luminosity of $1.1^{+0.6}_{-0.5}\times 10^{40}$ erg~s$^{-1}$. Likewise, for the two sources in J0842+1150, we find luminosities of $1.2^{+0.4}_{-0.3}\times 10^{40}$ erg~s$^{-1}$ for X-1 and $7^{+4}_{-3}\times 10^{39}$ erg~s$^{-1}$ for X-2. Due to the distances to these galaxies, the X-ray emission appears as a point source (see Figure~\ref{fig:xraySurfBright}) where the approximate $1\arcsec$ radius of the 90~per~cent PSFs correspond to $550-600$~parsecs. If we assume that a majority of this X-ray emission comes from either a single, dominant source or up to a few bright sources in each case, then these sources would qualify as ultraluminous X-ray sources (ULX). ULX are typically defined as non-nuclear, point-like, extragalactic sources with X-ray luminosities in excess of the Eddington limit for a 20~$M_\odot$ black holes (i.e., $\gtrsim 3\times 10^{39}$~erg~s$^{-1}$). 

For SHOC~486 the X-ray emission is localized to the optically bright, central region of the galaxy. 
However, SHOC~486 does not show any clear signatures of an AGN (i.e., optical line ratios, strong 1.4 GHz radio emission) and is likely a bright X-ray binary or collection of X-ray binaries. 
For J0842+1150, there is no obvious central region of the galaxy given its irregular morphology. Therefore, the two X-ray sources could be ULXs. The current data available for these two galaxies does not allow us to determine if any of these X-ray sources are dominated by emission from a single object. Multiple observations would be needed to look for any dramatic flux variations, which would indicate single, dominant sources. 

\begin{table*}
\centering
\begin{minipage}{110mm}
\caption{Spectral Fitting}\label{tab:spectral}
\begin{tabular}{lcccc}
\hline
Source     & Net Counts             & $\Gamma$      & Unabsorbed Flux                               & $L_X$ \\
           &  &               & $(10^{-15} \text{erg cm}^{-2}\text{ s}^{-1})$ & $(10^{39} \text{erg s}^{-1})$ \\ \hline \hline
SHOC 486   & $16.6\pm5.2$               & $1.3^{+1.3}_{-1.0}$ & $7.8^{+5.0}_{-3.8}$     & $14.3^{+9.1}_{-7.0}$ \\
X-1        & $11.6\pm3.5$               & $-$ & $-$     & $-$ \\
J0842+1150 & $81.8\pm10.4$              & $1.6\pm0.4$ & $17.5^{+3.8}_{-3.3}$   & $37.1^{+8.1}_{-7.0}$ \\
X-1        & $32.3\pm5.9$               & $1.4\pm0.5$ & $8.0^{+2.6}_{-2.2}$     & $17.0^{+5.5}_{-4.7}$   \\
X-2        & $12.7\pm3.8$               & $1.8^{+0.9}_{-0.7}$ & $3.5^{+1.7}_{-1.3}$     & $7.4^{+3.6}_{-2.8}$   \\ \hline
\end{tabular}
\\
\raggedright\textbf{Notes.} Using \texttt{XSPEC} to fit extracted spectra, we report the best-fit values for the absorbed power law model (\texttt{wabs*pow}). We find that all values are consistent with the assumed and calculated values from the photometric measures in Table~\ref{tab:results}. All values are for the $0.5-8$~keV energy range.\\
\end{minipage}
\end{table*}

\section*{ACKNOWLEDGEMENTS}
We thank the anonymous referee for helpful comments and suggestions that greatly improved the manuscript.
M.B. thanks Hai Fu for insightful discussions which helped to improve the quality of the paper.
The scientific results reported in this article are based on joint observations made by the Chandra X-ray Observatory and the NASA/ESA Hubble Space Telescope. Support for this work was provided by the National Aeronautics and Space Administration through Chandra Award Number G05-16048X issued by the Chandra X-ray Observatory Center, which is operated by the Smithsonian Astrophysical Observatory for and on behalf of the National Aeronautics Space Administration under contract NAS8-03060. Based on observations made with the NASA/ESA Hubble Space Telescope, obtained at the Space Telescope Science Institute, which is operated by the Association of Universities for Research in Astronomy, Inc., under NASA contract NAS 5-26555. These observations are associated with program \#13940. Support for program \#13940 was provided by NASA through a grant from the Space Telescope Science Institute, which is operated by the Association of Universities for Research in Astronomy, Inc., under NASA contract NAS 5-26555.

\bibliographystyle{mnras}
\bibliography{MyRefs2}

\end{document}